\documentclass[%
 reprint,
superscriptaddress,
 amsmath,amssymb,
aps,
]{revtex4-2}

\usepackage{graphicx}
\usepackage{dcolumn}
\usepackage{bm}
\usepackage{multirow}
\usepackage{makecell}
 \usepackage{float}
\usepackage{tabularx}

\usepackage{units}
\usepackage{color}
\usepackage{url}
\usepackage{soul}
\usepackage{upgreek}

\usepackage[colorlinks]{hyperref}
\hypersetup{%
	plainpages=true,
	breaklinks=true,
	hypertexnames=false,
	pageanchor=true,
	colorlinks=true,
	linkcolor={blue},
	citecolor={red},
	urlcolor={blue},
	anchorcolor={black}
}

\usepackage{mleftright} 

\newcommand{\figref}[1]{\mbox{Fig.~\ref{#1}}}
\newcommand{\figpanel}[2]{Fig.~\hyperref[#1]{\ref*{#1}(#2)}}
\newcommand{\figpanels}[3]{Fig.~\hyperref[#1]{\ref*{#1}(#2)-(#3)}}
\newcommand{\figpanelNoPrefix}[2]{\hyperref[#1]{\ref*{#1}(#2)}}

\renewcommand{\eqref}[1]{\mbox{Eq.~(\ref{#1})}}
\newcommand{\tabref}[1]{\mbox{Table~\ref{#1}}}

\newcommand{\bra}[1]{\mleft\langle #1 \mright |}
\newcommand{\ket}[1]{\mleft|#1 \mright \rangle}

\newcommand{\ketbra}[2]{\mleft| #1 \rangle \langle #2 \mright|}
\begin{document}

\preprint{APS/123-QED}

\title{Quantum process tomography of continuous-variable gates using coherent states}

\author{Mikael Kervinen}
\affiliation{Department of Microtechnology and Nanoscience, Chalmers University of Technology, 412 96 Gothenburg, Sweden}

\author{Shahnawaz Ahmed}
\affiliation{Department of Microtechnology and Nanoscience, Chalmers University of Technology, 412 96 Gothenburg, Sweden}

\author{Marina Kudra}
\affiliation{Department of Microtechnology and Nanoscience, Chalmers University of Technology, 412 96 Gothenburg, Sweden}

\author{Axel Eriksson}
\affiliation{Department of Microtechnology and Nanoscience, Chalmers University of Technology, 412 96 Gothenburg, Sweden}

\author{Fernando Quijandr\'{\i}a}
\altaffiliation[Present address: ]{Quantum Machines Unit, Okinawa Institute of Science and Technology Graduate University, Onna-son, Okinawa 904-0495, Japan.}
\affiliation{Department of Microtechnology and Nanoscience, Chalmers University of Technology, 412 96 Gothenburg, Sweden}


\author{Anton Frisk Kockum}
\affiliation{Department of Microtechnology and Nanoscience, Chalmers University of Technology, 412 96 Gothenburg, Sweden}

\author{Per Delsing}
\email{per.delsing@chalmers.se}
\affiliation{Department of Microtechnology and Nanoscience, Chalmers University of Technology, 412 96 Gothenburg, Sweden}

\author{Simone Gasparinetti}
\email{simoneg@chalmers.se}
\affiliation{Department of Microtechnology and Nanoscience, Chalmers University of Technology, 412 96 Gothenburg, Sweden}

\date{\today}

\begin{abstract}
Encoding quantum information into superpositions of multiple Fock states of a harmonic oscillator can provide protection against errors, but it comes with the cost of requiring more complex quantum gates that need to address multiple Fock states simultaneously. Therefore, characterizing the quantum process fidelity of these gates also becomes more challenging. Here, we demonstrate the use of coherent-state quantum process tomography (csQPT) for a bosonic-mode superconducting circuit. CsQPT uses coherent states as input probes for the quantum process in order to completely characterize the quantum operation for an arbitrary input state. We show results for this method by characterizing a logical quantum gate constructed using displacement and SNAP operations on an encoded qubit. With csQPT, we are able to reconstruct the Kraus operators for the larger Hilbert space rather than being limited to the logical subspace. This allows for a more accurate determination of the different error mechanisms that lead to the gate infidelity.

\end{abstract}


\maketitle


As quantum computing architectures become more refined, there is a surge of demand for methods to accurately characterize quantum gates and the processes limiting their performance. To meet this demand, methods like quantum process tomography~\cite{nielsen2002quantum}, randomized benchmarking~\cite{magesan2011scalable}, and gate-set tomography~\cite{greenbaum2015introduction} have been developed and tested on single-qubit \cite{chow2009randomized}, two-qubit \cite{mckay2016universal}, and three-qubit gates \cite{warren2022extensive}. At the same time, an alternative approach to quantum computing relies on encoding quantum information in harmonic oscillators, also referred to as continuous-variable systems, or bosonic modes \cite{joshi2021quantum}.
In contrast to two-level systems, the large Hilbert space of harmonic oscillators leaves a freedom in the choice of states in which to encode the information. This freedom can be exploited to render the encoded information robust against photon loss, the dominant source of error in oscillators, opening the door to hardware-efficient quantum error correction~\cite{terhal2020towards} and error-transparent or fault-tolerant gates~\cite{rosenblum2018fault,reinhold2020error,ma2020error}.

Compared to the two-level paradigm, the development of high-fidelity quantum gates and their characterization in bosonic modes is still at an early stage~\cite{cai2021bosonic,ma2021quantum}. In state-of-the-art experiments with superconducting circuits, quantum gates in bosonic modes are characterized by sandwiching the gate under test by encoding and decoding operations~\cite{heeres2017implementing}. These operations establish a mapping between the logical states of the qubit encoded in the oscillator and the states of an ancillary qubit, which can be more easily manipulated and read out. However, this technique suffers from serious limitations. First, the encoding and decoding operations are complex entangling gates whose fidelities are not guaranteed to exceed that of the gate under test, leading to significant state preparation and measurement (SPAM) errors. In addition, the decoding operation, followed by read-out of the ancillary qubit, reduces the full Hilbert space of the oscillator to that of a two-level system. This reduction hinders the possibility to reliably distinguish between different types of errors, as well as to characterize leakage errors~\cite{wallman2016robust}.

Here, we experimentally demonstrate the use of coherent-state quantum process tomography (csQPT) to characterize a quantum gate in a continuous-variable system. The gate acts on a logical qubit encoded in a bosonic mode, which is hosted by a microwave cavity in a superconducting circuit architecture.
Typically in harmonic oscillators, coherent states are the simplest states to prepare. By letting the gate under test act on a set of coherent-state probes and measuring the final states by direct Wigner tomography, we completely characterize the process in the Fock space of the oscillator. 

CsQPT was proposed and first implemented for quantum optical processes using homodyne tomography followed by maximum likelihood reconstruction~\cite{lobino2008complete, rahimi2011quantum}. 
To decrease the measurement overhead, we improve on
maximum-likelihood csQPT~\cite{anis2012maximum} by augmenting csQPT with a gradient-descent based learning algorithm using the idea of manifold learning~\cite{ahmed2022gradient}.
This method allows us to reconstruct quantum process matrices from a reduced number of data points, avoiding full state tomography. From the process matrices, we can generate any representation of the process, determine the process fidelity of the gate, and characterize the leakage outside of the computational subspace. Our approach goes beyond encoding-decoding schemes in two ways: it reduces SPAM errors by obviating the need for complex encoding and decoding operations, and it unlocks the access to the larger Hilbert space of the continuous-variable system, making it possible to correctly identify leakage errors.


%

Continuous-variable systems realized in particular with superconducting three-dimensional (3D) cavities coupled to an ancilla qubit have shown an unprecedented level of versatile control of quantum states~\cite{vlastakis2013deterministically, heeres2017implementing,campagne2020quantum, eickbusch2022fast, kudra2022robust}. We implement csQPT on a system consisting of a single-mode 3D superconducting $\lambda/4$-cavity~\cite{reagor2016, kudra2020high} and an ancilla transmon qubit~\cite{Koch2007}. The transmon qubit is used to provide a nonlinear control element which is necessary in order to control the harmonic energy levels of the cavity, as well as provide means to characterize the cavity states through direct Wigner tomography~\cite{vlastakis2013deterministically}. The transmon qubit with its own readout resonator is fabricated on a sapphire chip, and the chip is inserted into the cavity, where the qubit and the cavity are capacitively coupled. 

\begin{figure}[t]\centering
\begin{centering}
\includegraphics[width=1\columnwidth]{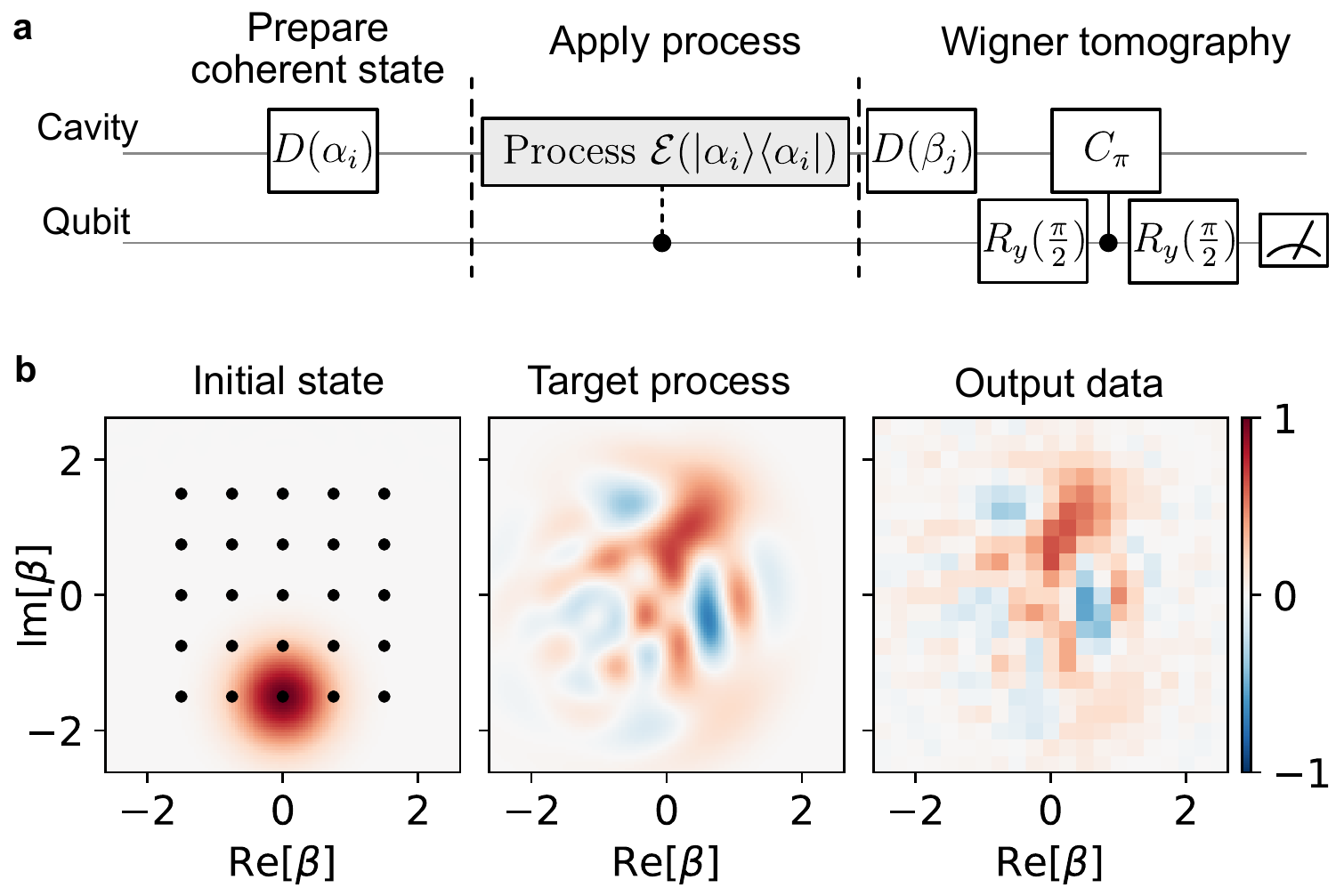}
\caption{\textbf{Protocol for the csQPT.} \textbf{a} Gate sequence for the process-tomography protocol. The probe states are coherent states in a 5 × 5 grid of complex displacements $\alpha_i$ (see the left plot in panel \textbf{b}). The process $\mathcal{E}$ is applied to each probe state. Finally, the cavity state is measured using Wigner tomography. The Wigner tomography consists of a cavity displacement D($\beta_j$), unconditional qubit $\tfrac{\pi}{2}$-pulse, conditional phase evolution $C_\pi$, and another qubit $\tfrac{\pi}{2}$-pulse followed by a readout. \textbf{b} Wigner functions at each step of the protocol. (Middle) shows the ideal target result of applying the process $\mathcal{E}$ to the coherent-state probe visualized in (Left). (Right) The data points are the values of the Wigner function in a 21 × 21 grid of complex-displacements $\beta_j$. The Wigner-function measurements are repeated for each input probe.}
\label{fig:protocol}
 \end{centering}
\end{figure}

To perform csQPT, we run experimental sequences consisting of three steps [\figpanel{fig:protocol}{a}]. First, we prepare the cavity in a coherent state $\ket{\alpha_i}$ -- our input probe. We create coherent states by passive thermalization to the ground (vacuum) state $\ket{0}$ followed by a displacement operation $D(\alpha_i)$. Next, we apply the quantum process $\mathcal{E}$ that we intend to characterize to the input state. Finally, we measure a displaced parity operator with the assistance of the ancillary qubit. To do so, we apply a displacement $D(\beta_j)$ to the cavity and then measure its photon parity by performing a Ramsey measurement that maps the parity to the $\sigma_z$ axis of the qubit \cite{vlastakis2013deterministically}. Averaging over this sequence and varying $\beta_j$ to map different regions of the phase space gives a direct measurement of the Wigner function $W(\beta)$ of the coherent state $\ket{\alpha_i}$, after it has been acted upon by the gate. 
We repeat this procedure for a rectangular $5\times5$ grid array of coherent-state probes spanning from $-1.5-1.5i$ to $1.5+1.5i$ [\figpanel{fig:protocol}{b}]. The amplitude of the coherent-state probes determines the maximum photon number that is populated and thereby sets a limit on the size of the cavity Hilbert space in which we can reliably reconstruct the process. We find our choice of the probe amplitude to be sufficient to reconstruct the process up to Fock state $|5\rangle$~\cite{supplement}.

The reconstruction of a process representation for $\mathcal E$ is performed using a gradient-based optimization that learns the Kraus representation of the process~\cite{ahmed2022gradient}. The Kraus operators are learned by minimizing a loss function that is the squared error between the measured Wigner data points and the corresponding Wigner points predicted by the reconstructed Kraus operators. Our optimization procedure is constrained to the manifold of completely positive and trace preserving quantum operations with appropriate restrictions on the Kraus operators~\cite{supplement}.
%
%
Reconstructing the process at the Kraus level allows us to limit the size of the process representation by restricting the number of Kraus operators. We therefore limit our reconstruction to a low rank, so that we can learn the dominant process channels without having to reconstruct the full-rank process. Additionally, the noise in the data may not allow the reconstruction of all the loss channels even if more Kraus operators are introduced. With this method, we can reconstruct the Kraus operators directly without an intermediate step of reconstructing the density matrices of the output states. 
%

\begin{figure}[t]\centering
\begin{centering}
\includegraphics[
width=1\columnwidth]{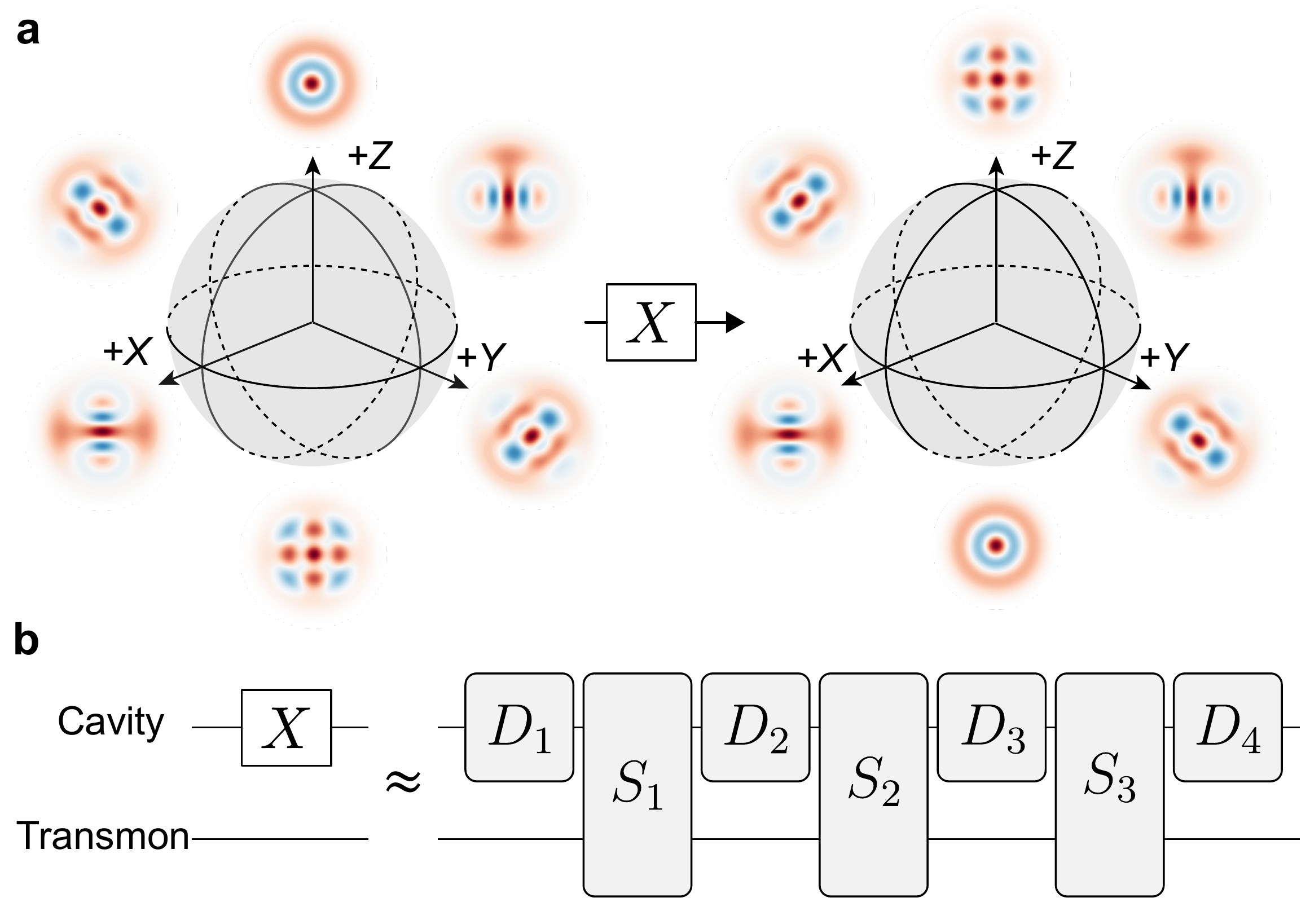}
\caption{\textbf{X-gate in binomial encoding.} \textbf{a} The gate is defined by mapping the six cardinal states in the encoded subspace to their respective targets. The gate is defined only for the encoded subspace, but we can characterize it generally by analyzing its action on the full Hilbert space and not just the cardinal states. \textbf{b} The X-gate for the cavity-encoded logical state is composed of three SNAP operations and four displacement operations.}
\label{fig:Fig2}
 \end{centering}
\end{figure}

We test csQPT on a quantum logical gate that swaps the population of the states $|0_L\rangle$ and $|1_L\rangle$, i.e., a logical X-gate (\figref{fig:Fig2}). We choose the binomial encoding
\begin{align*}
&|0_L\rangle = |2\rangle, \\
&|1_L\rangle = \tfrac{1}{\sqrt{2}}(|0\rangle+|4\rangle),
\end{align*}
which is the lowest-order binomial code that can be corrected for the single-photon loss error in the cavity~\cite{hu2019quantum}. We implement the gate as a series of displacement $D(\alpha)$ and Selective Number-dependent Arbitrary Phase (SNAP) operations $S(\theta)$~\cite{heeres2015cavity}. These two operations provide universal control of the cavity states~\cite{krastanov2015universal}. We numerically optimize the gate sequence~\cite{fosel2020efficient} and the pulse envelopes following the method described in Ref.~\cite{kudra2022robust}, which was shown to reduce the gate length while making the pulses more robust against variations of the system parameters. In SNAP gates, the qubit starts and ends in the ground state with high probability regardless of the initial state of the cavity, so that the same qubit becomes available for the Wigner-tomography measurement protocol. Previously, SNAP and displacement operations had only been used in the context of state preparation. Here, we show that these operations can also be used to efficiently implement logical gates. In particular, we implement the X-gate with only three SNAP gates and 4 displacements. Compared to gate implementations based on fully numerical optimal control, in which the transmon-cavity composite system is driven simultaneously~\cite{heeres2017implementing,hu2019quantum}, separate gates on the cavity and on the transmon are more easily parameterized, and the effects of the individual operations are more transparent~\cite{kudra2022robust}. 

\begin{figure}[t]\centering
\begin{centering}
\includegraphics[
width=1\columnwidth]{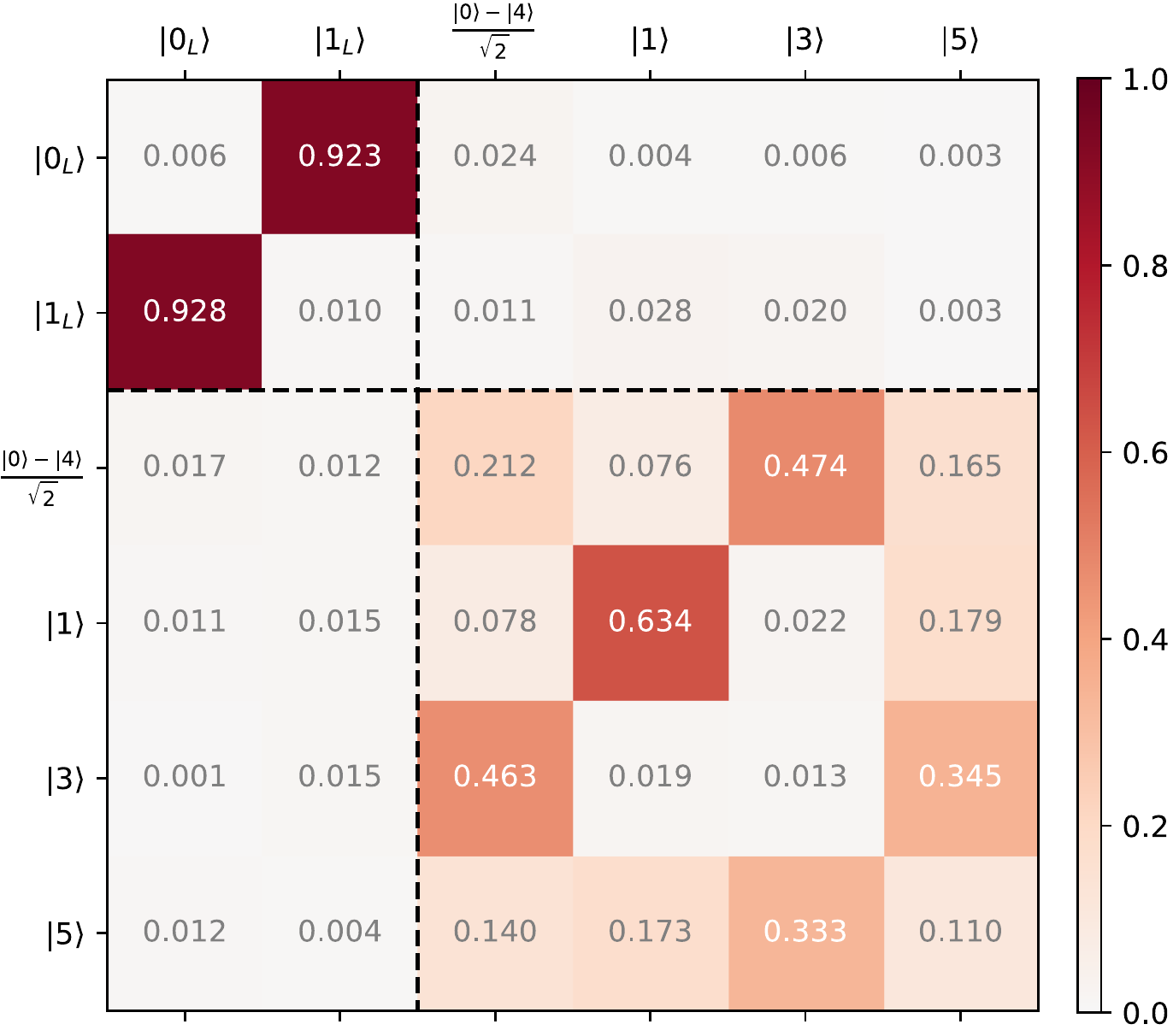}
\caption{\textbf{Population transfer matrix.} The upper-left block represents the logical subspace, where we can identify the X-gate that swaps states  $|0_L\rangle$ and $|1_L\rangle$. The columns correspond to the input states, while rows correspond to the output states.}
\label{fig:ptm}
 \end{centering}
\end{figure}

We characterize the X-gate by running the process tomography sequence and process reconstruction described above. From the obtained Kraus operators, we construct the population transfer matrix (\figref{fig:ptm}). The elements of the matrix describe the population distribution of the final state in a chosen basis, given one of the basis states was prepared as the initial state. Instead of the usual Fock basis, the matrix is presented in a basis given by $\{\ket{0_L}, \ket{1_L}, \tfrac{1}{\sqrt{2}}(|0\rangle-|4\rangle), \ket{1}, \ket{3}, \ket{5}\}$, which transparently shows the effect on the logical basis vectors. 
In this representation, we identify the X-gate in the logical subspace in the upper-left block of the population matrix. By inspecting the matrix entries we see that the swap operation between $|0_L\rangle$ and $|1_L\rangle$ is successful, with a population transfer between \unit[92]{\%} and \unit[93]{\%} and a population loss on the order of \unit[7]{\%}. The elements underneath the logical states describe population leakage outside of the computational subspace. For example, when preparing $|0_L\rangle$, we observe that most of the population loss is due to leakage to states outside of the computational subspace. The largest leakage is into $\tfrac{1}{\sqrt{2}}(|0\rangle-|4\rangle)$, which is one of the no-jump evolution error states of the binomial code~\cite{girvin2021introduction}. Similarly, the largest leakage from $|1_L\rangle$ is into $\ket{1}$ and $\ket{3}$, which are the main error states of the binomial code.

\begin{figure}[t]\centering
\begin{centering}
\includegraphics[
width=1\columnwidth]{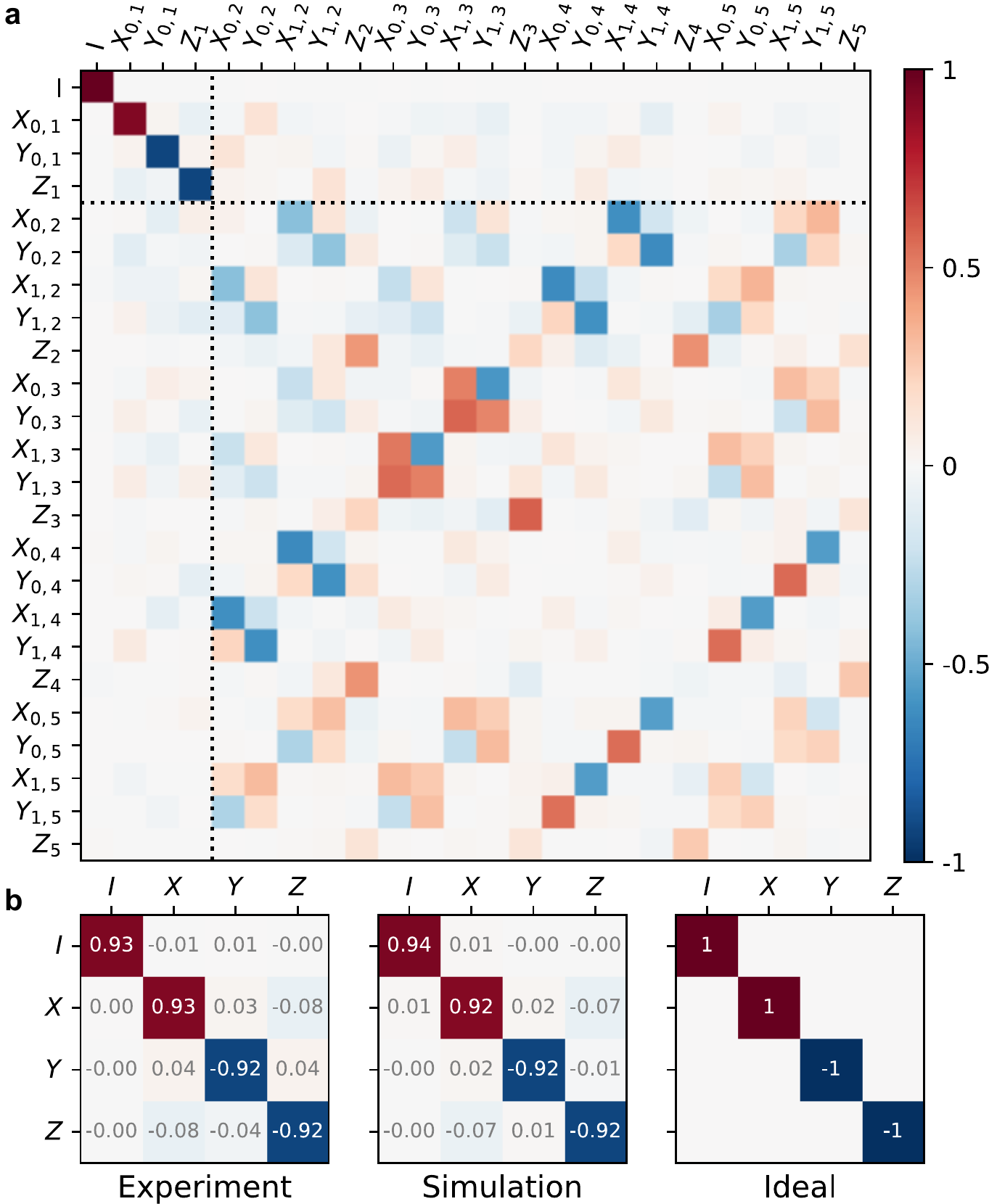}
\caption{\textbf{Gell-Mann transfer matrix of an X-gate.} \textbf{a} The upper-left block represents a Pauli-transfer-matrix-like structure for the logical subspace. The other elements in the Gell-Mann transfer matrix represent couplings between different levels of the system. They are given in a basis defined by the Gell-Mann matrices. \textbf{b} Comparison of the experimental (reconstructed), simulated, and ideal two-level Pauli transfer matrices that are obtained by restricting the Gell-Mann transfer matrix to the logical subspace.}
\label{fig:gtm}
 \end{centering}
\end{figure}
The population transfer matrix
does not describe the coherence between the chosen basis states. As such, it only offers a partial representation of the quantum process. To provide a complete description of the process $\mathcal E$, we use a generalization of the Pauli transfer matrix from two-level systems to $d$-level systems. We refer to this generalization as the Gell-Mann transfer matrix~\cite{supplement}. We show the transfer matrix in \figref{fig:gtm}, where we have only included the elements that couple to the logical states up to a Fock state $\ket{5}$. In the upper-left corner of the Gell-Mann transfer matrix, we can identify a Pauli-transfer-matrix-like block of an X-gate for the two-level logical subsystem. In \figref{fig:gtm}{b}, we present the experimental Pauli transfer matrix that is calculated from the Kraus operators alongside the simulated and ideal transfer matrices. Coherent errors within the logical subspace appear as off-diagonal elements in the Pauli transfer matrix for this particular gate. The detailed information of the leakage out of (into) the logical subspace is given by the off-diagonal blocks below (next to) the computational subspace~\cite{helsen2019spectral}.

We can now calculate the average gate fidelity~\cite{Nielsen2002} between the reconstructed process $\mathcal{E}$ and the targeted logical gate $U$ in the $d=2$-dimensional logical subspace, also considering leakage, as
\begin{equation}
\label{eq:avgf}
F_g(\mathcal{E}, U) = \frac{d F_{\text{pro}}(\mathcal E, U) + 1 - L_L(\mathcal E)}{d+1}.
\end{equation}
Here $F_{\text{pro}}(\mathcal E, U)$ is the process fidelity, which can be written down using any representation of the process $\mathcal{E}$, e.g., the Kraus operators~\cite{nielsen2002quantum}, the Choi matrix~\cite{Fletcher2007}, or the Pauli transfer matrix~\cite{chow2012universal}. The average leakage rate is defined as $L_L(\mathcal E) = \int d{\psi_L} \mathcal{E}(\ketbra{\psi_L}{\psi_L}) = L(\mathcal{E}(\frac{\mathbb I_L}{d}))$~\cite{Wood2018}, where $L(\rho) = 1 - \text{Tr}[\mathbb I_L \rho]$ quantifies the leakage from the logical subspace and the integral is over all  states in the logical subspace. The unitary $U = \ketbra{0_L}{1_L} + \ketbra{1_L}{0_L}$ represents the logical $X$-gate and the projector $\mathbb I_L = \ketbra{0_L}{0_L} + \ketbra{1_L}{1_L}$ is the identity in the logical subspace. We can write \eqref{eq:avgf} using the Kraus representation of the process in the full Hilbert space (see Methods), providing us information about the leakage out from the computational subspace. The average population leakage out of the computational subspace is calculated to be \unit[6.6]{\%}.
The average gate fidelity obtained from the reconstructed process is $F_g(\mathcal E, U) = \unit[92.8]{\%}$. In order to estimate the effects of decoherence with known parameters of our system, we calculate the average gate fidelity from Lindblad master-equation simulations, with our measured decoherence rates. We obtain an expected gate fidelity of \unit[92.9]{\%} from the simulations, which is in close agreement with the average gate fidelity given by our method. We thus show the first experimental demonstration that relatively high-fidelity logical quantum gates can be composed of displacement and SNAP operations.

With continuous-variable gates, the leakage into energy levels outside of the computational basis presents a problem that needs to be carefully addressed. With our method, we avoid a large component in the error estimation that comes from estimating the dimension of the Hilbert space that the errors leak into~\cite{heeres2017implementing}. We foresee that with further analysis of the Gell-Mann transfer matrix, more detailed error analysis can be performed. In particular, having access to the process-matrix elements beyond the logical subspace opens up possibilities to study different error models in order to more accurately pinpoint the origin of the gate infidelity~\cite{helsen2019spectral}. Our results establish the use of csQPT to understand the error mechanisms affecting continuous-variable quantum gates, paving the way towards better-performing bosonic codes. Our method can also be directly applied to simultaneously characterize the effect of a certain operation on both the code subspace and the error subspace, which could assist the design of quantum error correction sequences.



\begin{acknowledgments}
We would like to thank Mats Myremark and Lars J\"onsson for machining the cavity, and Amr Osman for fabricating a similar qubit chip. The simulations and visualization of the quantum states were performed using QuTiP~\cite{Johansson2012,Johansson2013}, NumPy~\cite{Harris2020array}, and Matplotlib~\cite{Hunter2007}. The automatic differentiation tool Jax~\cite{Jax2018} was used for process reconstruction and the optimization of gate parameters. This work was supported by the Knut and Alice Wallenberg foundation via the Wallenberg Centre for Quantum Technology (WACQT) and by the Swedish Research Council. The chips were fabricated at the Chalmers Myfab cleanroom. We acknowledge IARPA and Lincoln Labs for providing the TWPA used in this experiment.

\end{acknowledgments}

\bibliography{apssamp}

\section*{Methods}
\label{sec:methods}
\subsection*{Kraus representation of a quantum process}

In general, a quantum process is described by a completely positive trace-preserving map $\mathcal E$ between the input and output states. This map can be represented with Kraus operators $K_i$ as
\begin{equation}
    \rho' = \mathcal E(\rho) = \sum_i^r K_i \rho K_i^\dagger.
\end{equation}
The number $r$ of Kraus operators can take values from $1$, which corresponds to a unitary process, up to $d^2$ in a Hilbert space with dimension $d$. The Kraus representation ensures the complete positivity of the process. Trace preservation is imposed by satisfying the constraint
\begin{equation}
    \sum_i^r K_{i}^\dagger K_i = \mathbb{I}.
\end{equation} 

\subsection*{Gell-Mann transfer matrix}
Similar to the $(d=2)$ Pauli transfer matrix~\cite{chow2012universal}, we define the Gell-Mann transfer matrix of a process $\mathcal E$ as a single real-valued matrix $\Lambda^{\mathcal E}$, whose elements are bounded between $-1$ and $+1$. The Gell-Mann matrices $\{G_i\}$ \cite{bertlmann2008bloch}, including the identity matrix, form an operator basis for the symmetry group $SU(d)$ similar to how the Pauli matrices generate arbitrary unitaries in $SU(2)$. The Gell-Mann transfer matrix is computed by applying the quantum process to the Gell-Mann matrices and then computing the matrix elements $\Lambda^{\mathcal E}_{ij}$ as
\begin{equation}
    \Lambda^{\mathcal{E}}_{ij} = \frac{1}{d}\text{Tr}[G_j \mathcal E(G_i)].
    \label{eq:gtm}
\end{equation}
We can then write the action of the process on some quantum state $\rho$ using the Gell-Mann transfer matrix, by vectorizing $\rho$ using the operator basis consisting of the Gell-Mann matrices, as $\mathcal{E}(\rho) = \Lambda^{\mathcal{E}} \vec \rho$ with the elements of the vectorized density matrix given by ${(\vec{\rho})}_i = \text{Tr}[G_i\rho]$.
\subsection*{Average gate fidelity for the logical subspace}
\label{subsec:avggate}
The average gate fidelity between a unitary operation $U$ representing a quantum gate and a process $\mathcal E$ can be defined as~\cite{Nielsen2002}
\begin{equation}
\label{eq:avgfstart}
    F_g(\mathcal{E}, U) = \int d \ket{\psi} \bra{\psi} U^{\dagger} \mathcal {E}(\ketbra{\psi}{\psi})U\ket{\psi},
\end{equation}
where the integral is over the set of all input states $\ket{\psi} = \sum_n c_n \ket{n}$ with $c_n \in \mathbb C$ and $\sum_n |c_n|^2 = 1$. Using the Kraus representation for the process $\mathcal E$ we can generally write the integral as
\begin{eqnarray}
\label{eq:avgfexpanded}
    F_g(\mathcal{E}, U) &=& \sum_i \int d \ket{\psi} \bra{\psi} U^{\dagger} K_i \ketbra{\psi}{\psi}K_i^{\dagger}U\ket{\psi} \nonumber \\
    &=& \sum_i \sum_{klmn} \bra{k}U^{\dagger} K_i\ket{l}\bra{m}K_i^{\dagger}U\ket{n} \int d \ket{\psi} c_k^* c_l c_m^* c_n \nonumber \\
    &=& \sum_i \sum_{klmn} \bra{k}U^{\dagger} K_i\ket{l}\bra{m}K_i^{\dagger}U\ket{n}\frac{(\delta_{kl} \delta_{mn} + \delta_{kn} \delta_{lm})}{d(d+1)}  \nonumber \\
\end{eqnarray}
where the value of the integral $\int d \ket{\psi} c_k^* c_l c_m^* c_n = \frac{(\delta_{kl} \delta_{mn} + \delta_{kn} \delta_{lm})}{d(d+1)}$ over the set of input states. This integral can be evaluated in several ways~\cite{Willsch2020}; one of the simplest is to reduce it to a set of Gaussian integrals~\cite{Jin2021}. In order to obtain a simple analytical expression for $F_g(\mathcal E, U)$, we can identify the various terms as taking the trace in the logical subspace such that
\begin{eqnarray}
\label{eq:avgfderived}
   F_g(\mathcal{E}, U)  &=& \sum_i  \sum_{km} \bra{k}U^{\dagger} K_i\ket{k}\bra{m}K_i^{\dagger}U\ket{m} \frac{1}{d(d+1)} + \nonumber \\
    && \sum_i \sum_{kl} \bra{k}U^{\dagger} K_i\ket{l}\bra{l}K_i^{\dagger}U\ket{k} \frac{1}{d(d+1)} \nonumber \\
    &=& \sum_i  \frac{ |\text{Tr} [U^{\dagger} K_i]|^2 + \text{Tr}[U^{\dagger} K_i (\sum_{l} \ket{l}\bra{l})K_i^{\dagger}U]}{d(d+1)}  \nonumber \\
     &=& \frac{\sum_i|\text{Tr} [U^{\dagger} K_i]|^2 + \text{Tr}[U^{\dagger} \mathcal{E}(\mathbb I_L)U]}{d(d+1)} \nonumber \\
     &=& \frac{\sum_i|\text{Tr} [U^{\dagger} K_i]|^2 + d \text{Tr}[\mathbb I_L \mathcal{E}(\frac{\mathbb I_L}{d})]}{d(d+1)}.
\end{eqnarray}
Since we are interested in the average gate fidelity within the logical subspace, we only consider states in the logical basis as $\ket{\psi} = \ket{\psi_L} = c_0 \ket{0_L} + c_1 \ket{1_L} $ such that the identity operation is given by $\mathbb I_L = \sum_l \ketbra{l}{l} = \ketbra{0_L}{0_L} +  \ketbra{1_L}{1_L} $. We have used the property that taking the trace is invariant to a basis transformation such that we can write $U$ and $K_i$ in any basis. We have also used the cyclic property of the trace to write $\text{Tr}[U^{\dagger}\mathcal{E}(\mathbb I_L)U] = \text{Tr}[U^{\dagger}U\mathcal{E}(\mathbb I_L)] = d \text{Tr}[\mathbb {I}_L \mathcal{E}(\frac{\mathbb I_L}{d})]$ in the last step.

We can easily reduce \eqref{eq:avgfderived} to \eqref{eq:avgf} from Ref.~\cite{Wood2018} by identifying the process fidelity in terms of Kraus operators as~\cite{nielsen2002quantum}
\begin{equation}
F_{\text{pro}}(\mathcal E, U) = \frac{\sum_i |\text{Tr} [U^{\dagger} K_i]|^2}{d^2},
\end{equation}
such that \eqref{eq:avgfderived} can be written as
\begin{eqnarray}
F_g(\mathcal{E}, U) &=& \frac{d^2 F_{\text{pro}}(\mathcal E, U) + d \text{Tr}[\mathbb I_L \mathcal{E}(\frac{\mathbb I_L}{d})]}{d(d+1)} \nonumber \\
&=& \frac{d F_{\text{pro}}(\mathcal E, U) + 1 - L_{L}(\mathcal{E})}{d+1}.
\end{eqnarray}

\onecolumngrid
\pagebreak
\setcounter{table}{0}
\renewcommand{\thetable}{S\arabic{table}}
\setcounter{figure}{0}
\renewcommand{\thefigure}{S\arabic{figure}}

\setcounter{section}{0}
\renewcommand{\thesection}{S\arabic{section}}
\setcounter{equation}{0}
\renewcommand{\theequation}{S\arabic{equation}}

\setcounter{page}{1}

\section{System parameters}
\label{app:A}

The experimental setup is the same as in Ref.~\cite{kudra2022robust}. The qubit-cavity system in the dispersive regime is described by the Hamiltonian
\begin{align}
\label{eq:hamil}
    H = \omega_c a^\dagger a + \dfrac{K_c}{2}(a^\dagger)^2 a^2 + \omega_{q} b^\dagger b 
    + \chi a^\dagger a b^\dagger b + \dfrac{\chi'}{2}(a^\dagger)^2 a^2 b^\dagger b,
\end{align}
where $\omega_{c,q}$ are the resonance frequencies of the cavity and the transmon qubit, respectively; $\chi$ is the dispersive shift between the two modes; $\chi'$ is the higher-order dispersive shift; $K_c$ is the self-Kerr of the cavity; $a^\dagger$ ($a$) is the creation (annihilation) operator of the cavity field, and, similarly, $b^\dagger$ ($b$) is the raising (lowering) operator of the qubit. The experimentally measured system parameters are listed in \tabref{tab:Hamilton_param}.

\begin{table}[h]
\caption{\label{tab:Hamilton_param}
Parameter values for the system Hamiltonian and the measured coherence times. }
\begin{ruledtabular}
\begin{tabular}{ccc}
Parameter & Symbol & Value \\ 
\hline
Transmon frequency & $\omega_q/2\pi$ & \unit[$6.281$]{GHz} \\
Cavity frequency & $\omega_c/2\pi$ & \unit[$4.454$]{GHz} \\ 
Readout resonator frequency & $\omega_r/2\pi$ & \unit[$7.348$]{GHz} \\
Transmon-cavity cross-Kerr & $\chi_{qc}/2\pi$& \unit[$-2.22$]{MHz}\\
Transmon-resonator cross-Kerr & $\chi_{qr}/2\pi$& \unit[$-1.4$]{MHz} \\
Cavity self-Kerr & $K_c/2\pi$& \unit[$-2.7$]{kHz}  \\
Transmon anharmonicity & $\alpha_{q}/2\pi$& \unit[$-353$]{MHz} \\
Transmon-cavity higher-order cross-Kerr. & $\chi'_{qc}/2\pi$& \unit[$-14$]{kHz}\\
Transmon energy decay time & $T_{1q}$ & \unit[$\approx 38$]{$\upmu$s} \\
Transmon Ramsey & $T_{2 q}$ & \unit[$\approx 47$]{$\upmu$s} \\
Cavity decay time & $T_{1c}$ & \unit[$\approx 315$]{$\upmu$s} \\
Cavity Ramsey time & $T_{2c}$ & \unit[$\approx 478$]{$\upmu$s} \\
Transmon thermal population & $\bar{n}_q$ & 0.005 \\
Cavity thermal population & $\bar{n}_c$ & 0.003 \\
\end{tabular}
\end{ruledtabular}
\end{table}

\section{Wigner tomography}
Since SNAP gates return the qubit to the ground state with high probability, the qubit is immediately available for Wigner-function measurement. We measure the Wigner function by first displacing the state with a short resonant pulse on the cavity with varying phase. Next, using an unconditional $\pi/2$-pulse followed by a waiting time $1/(2\chi)$ we output a $\pi/2$ or a $-\pi/2$ pulse. Each point is averaged $1000$ times. The two readout amplitudes are subtracted to obtain the Wigner amplitude $A(\beta)$. The Wigner function is normalized such that the resulting state represents a physical state with $\mathrm{Tr}(\rho)=1$ in order to minimize the effects of the limited bandwidth $\pi/2$ pulses and the measurement infidelity. 

\begin{figure}[h]
    \centering
    \includegraphics[width=0.6\columnwidth]{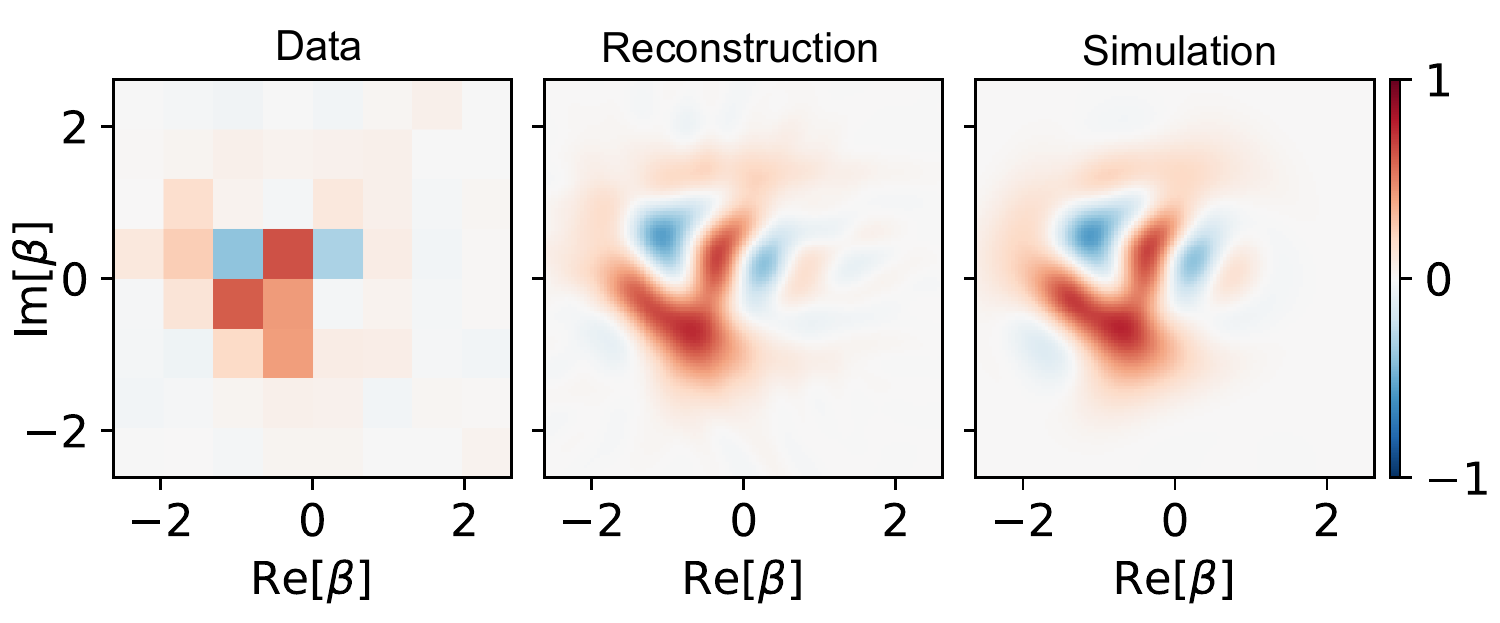}
    \caption{\textbf{Wigner tomography with a small number of points.} The process reconstruction is able to find the underlying process even from a very coarse data, showing that we only need a few measurements for each probe in our reconstruction.}
    \label{fig:wigner_points}
\end{figure}

For the Wigner tomography, we use a square grid of displacements with maximum amplitude on the real and imaginary axis of the phase space $\operatorname{Re}(\beta_{max}) = 2.62$. Since the gate that we aim to benchmark is photon-number-preserving, the features of the Wigner function are well captured within the chosen maximum displacement. The required number of Wigner displacements can be optimized numerically by minimizing the condition number that tells how many measurements are needed to reconstruct the underlying density matrix with a certain amount of photons~\cite{Reinhold_PhD}. However, since we do not need an intermediate step of finding the underlying density matrices, the number of Wigner displacements can be significantly reduced. Our process reconstruction is able to find the underlying process even from very coarse data (\figref{fig:wigner_points}). Further optimization could be done by numerically finding the optimal set of Wigner displacements instead of the grid pattern.

\section{Logical gate optimization}
We demonstrate a single quantum gate on the logical qubit, namely, the X-gate. Different gates can also be realized on the binomial encoding, leading to a universal gate set~\cite{hu2019quantum}. The logical gate is realized with a set of displacement and SNAP operations. The SNAP gate can be mathematically described as~\cite{fosel2020efficient}:
\begin{align}
\label{eq:SNAP}
    S(\vec{\theta}) = \sum_{n=0}^{N}e^{i\theta_{n}}\ket{n}\bra{n},
\end{align}
which amounts to applying the phase $\theta_{n}$ to the $n$th Fock state.

We calculate the displacement amplitudes and SNAP angles by mapping the cardinal points on the Bloch sphere to their ideal respective points after the gate. We then use a gradient descent based optimization that finds the optimal gate parameters for the chosen number of SNAP operations. In total, the X-gate consists of 4 displacement operations and 3 SNAP operations. The ideal gate in the presence of zero loss has an average gate fidelity of \unit[99.4]{\%}.

\begin{table}[h]
\caption{\label{tab:dispSNAPpar}
Gate parameters used to create the X-gate.}
\begin{ruledtabular}
\begin{tabularx}{0.8\textwidth}{ccc}
Operation & Symbol & Values\\
\hline
Displacement & $\beta_1$, $\beta_2$, $\beta_3$, $\beta_4$ & 0.610, 0.612, -0.612, -0.610\\[4mm]
SNAP & $\vec{\theta}_1$&\makecell{(-0.67791071, -0.09477794, -1.38876256,  0.53945346,  0.31723896,\\
        -1.30273005,  0.10376766,  2.65894245, -1.10789012,  0.50023422)}\\[4mm]
 &$\vec{\theta}_2$&\makecell{(0.,  2.7514428 ,  1.55112927,  2.31904201, -1.11177419, \\
         1.06874247,  0.33546735, -0.44872477, -0.77601542, -0.73785501)}\\[4mm]
 &$\vec{\theta}_3$&\makecell{(0.45755119,  1.03469991, -0.22172176,  1.70482232,  1.49607879,\\
        -0.12840042,  1.27637479, -2.36464223,  0.,  1.66335354)}\\
\end{tabularx}
\end{ruledtabular}
\end{table}

The individual SNAP operations are chosen to be \unit[700]{ns} long $(t_\mathrm{SNAP} > \tfrac{2\pi}{\chi_{qc}})$, whereas the displacement operations are \unit[100]{ns} long, making the total logical gate length $\unit[2.5]{\mu s}$. The SNAP operations are numerically optimized with the Boulder Opal optimizer from Q-CTRL. The resulting pulse envelopes perform the SNAP gate and also correct for the Kerr evolution during the gate.

\section{Coherent states as probes}
\begin{figure}[h]
    \centering
    \includegraphics[width=0.5\columnwidth]{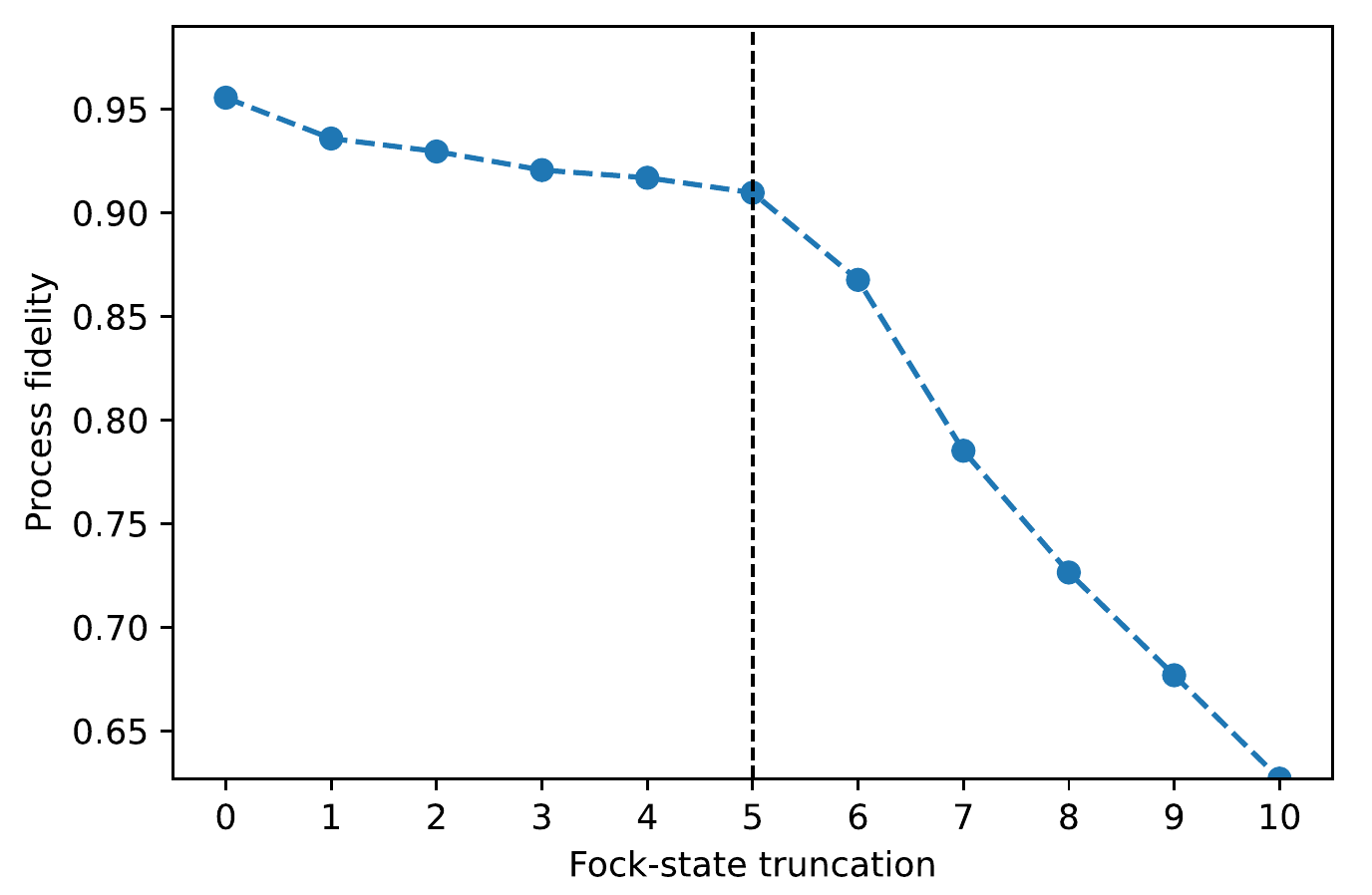}
    \caption{Process fidelity as a function of Fock-space truncation. The dashed line marks the Fock state limit up until which we can reliable gain information of our process.}
    \label{fig:n_cut}
\end{figure}

The coherent-state probes are generated by linearly driving the cavity with short resonant pulses. The probes are chosen in a grid pattern with the maximum amplitude along the real and imaginary axes set to $\alpha_{max}=1.5$. The choice of the $\alpha_{max}$ will define the maximum input Fock state that the process reconstruction is able to gain information about. In general, the amplitude of the probes and the number of probes required will depend on the process. For example, a process that is invariant with respect to the phase of an input coherent state can be probed with fixed-phase probes, only varying the amplitude. We report the process fidelity as a function of the Fock-state truncation in \figref{fig:n_cut}. The process fidelity is acquired by going from the Kraus representation to the Choi matrix representation \cite{wood2011tensor}. The Choi matrix can describe the complete process up to a certain Fock state truncation. The process fidelity is defined as
\begin{equation}
\label{eq:proF}
    F_\mathrm{pro} = F(\rho_\mathcal{E},\rho_\mathcal{F}),
\end{equation}
where $\rho_\mathcal{E}$ is the normalized Choi matrix of the process $\mathcal{E}$, $\rho_\mathcal{F}$ is the ideal process, and $F$ is the state fidelity.
After Fock state $\ket{5}$ we see a rapid decline in the process fidelity. Based on this, we can claim that the reconstruction of the gate dynamics is reliable only up to Fock state $|5\rangle$, which agrees with a simple calculation of the average maximum photon number $n_{max} = \mathrm{Re}(\alpha_{max})^2+\mathrm{Im}(\alpha_{max})^2 = 4.5$. Therefore, the selected probes are enough to probe the dominating effects for a logical gate spanned by basis states $|0_L\rangle = |2\rangle$ and $|1_L\rangle = \tfrac{1}{\sqrt{2}}(|0\rangle+|4\rangle)$. 

\section{Kraus reconstruction}
The number of Kraus operators was chosen to be 4, since adding more Kraus operators to the reconstruction did not have any effect on the fidelity. We reconstruct the Kraus operators directly without an intermediate step of reconstructing the density matrices by calculating the Wigner function from the predicted Kraus operators. The Kraus operators are reconstructed with a Hilbert space size $d=32$, since the Wigner functions are informationally very dense and can have effects that are explained by a population in high Fock states. The reconstruction is done by minimizing a loss function between the measured Wigner data and a reconstruction of the process. The loss function that the reconstruction tries to minimize is the squared distance L2-norm between the data and the simulated Wigner function:
\begin{equation}
    L2 = (y_\mathrm{data}-y_\mathrm{predicted})^2.
\end{equation}
\begin{figure}[h]
    \centering
    \includegraphics[width=0.5\columnwidth]{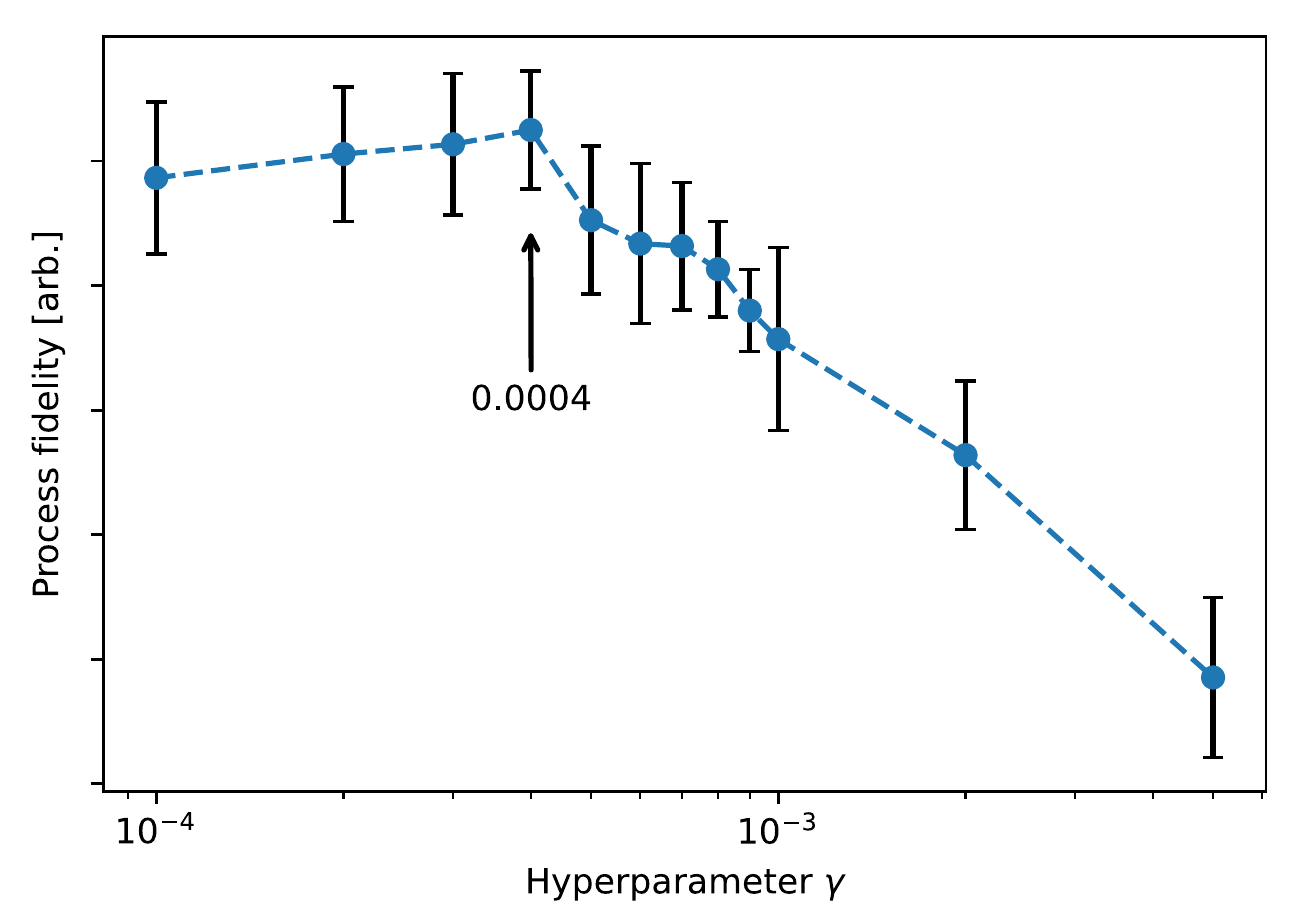}
    \caption{Hyperparameter optimization. Based on the result, we select a small value $\gamma = 0.0004$ for the L1 coefficient.}
    \label{fig:L1}
\end{figure}

In addition, we use an absolute error L1-loss function with a small coefficient $\gamma$:
\begin{equation}
    L1 = \gamma|\mathbb K|,
\end{equation}
where $|\mathbb K|$ represents the absolute value of the Kraus elements. The effect of the L1-loss regularization is to promote sparse Kraus operators by penalizing the reconstruction from having Kraus operators with too many non-zero elements. This helps to disregard noise in the measurement data. We find an appropriate value of the hyperparameter $\gamma$ by comparing the reconstructed Wigner functions to the simulated target state, which is obtained from master-equation simulation that includes independently measured loss rates of the system. We calculate the process fidelity between the Wigner functions of the simulated process and the reconstructed ones according to \eqref{eq:proF}, and select the hyperparameter value that maximises the process fidelity. However, maximizing the process fidelity between the simulation and the reconstruction of the process does not equal maximizing the fidelity of the logical gate operation. The role of the L1 parameter can be interpreted as follows: at small parameter values we reconstruct too much noise, whereas at very high parameter values the process is simplified too much. The impact of the noise in our process could be reduced by increasing the number of measurement averages but with the cost of measurement time. Alternatively, parameterized Kraus operators based on a loss model could be used to fit noisy data.


\section{Gell-Mann transfer matrix}
We introduce the Gell-Mann transfer matrix (GTM) as a generalization of the Pauli transfer matrix (PTM) for qubits. The PTM is a matrix representation of a general quantum process $\mathcal E$ evaluated by the action of $\mathcal{E}$ on the normalized set of $n$-qubit Pauli matrices $\{P_0, P_1, P_2, ..., P_{d^2 - 1}\}$ with $P_0 = \mathbb I$, and $d=2^n$ being the Hilbert space dimension. In the case of a single qubit, we have $d=2$ such that the matrices that form the operator-basis is given by $P_i = \{\mathbb I, \sigma_x, \sigma_y, \sigma_z\}$. We can write any density matrix using this basis as $\rho = \frac{\mathbb I}{d} + \sum^{d^2 - 1}_{i=1} b_i P_i$.
This set of matrices allows us to define the PTM $R$ with its matrix elements given by
\begin{equation}
    R_{ij} = \frac{1}{d}\text{Tr}[P_i \mathcal E(P_j)].
\end{equation}

We now consider an extension of the PTM to a single three-level system (a qutrit) that has a Hilbert-space dimension $d=3$. In this scenario, the Gell-Mann matrices extend the Pauli matrices from $SU(2)$ to $SU(3)$. The eight standard Gell-Mann matrices are defined for $d=3$ as
\begin{eqnarray}
    G_1 = \begin{pmatrix} 0 & 1 & 0 \\ 1 & 0 & 0 \\ 0 & 0 & 0 \end{pmatrix} \quad
G_2 = \begin{pmatrix} 0 & -i & 0 \\ i & 0 & 0 \\ 0 & 0 & 0 \end{pmatrix} \quad
G_3 &=& \begin{pmatrix} 1 & 0 & 0 \\ 0 & -1 & 0 \\ 0 & 0 & 0 \end{pmatrix} \quad
G_4 = \begin{pmatrix} 0 & 0 & 1 \\ 0 & 0 & 0 \\ 1 & 0 & 0 \end{pmatrix} \nonumber \\
G_5 = \begin{pmatrix} 0 & 0 & -i \\ 0 & 0 & 0 \\ i & 0 & 0 \end{pmatrix} \quad
G_6 = \begin{pmatrix} 0 & 0 & 0 \\ 0 & 0 & 1 \\ 0 & 1 & 0 \end{pmatrix} \quad
G_7 &=& \begin{pmatrix} 0 & 0 & 0 \\ 0 & 0 & -i \\ 0 & i & 0 \end{pmatrix} \quad
G_8 = \frac{1}{\sqrt{3}} \begin{pmatrix} 1 & 0 & 0 \\ 0 & 1 & 0 \\ 0 & 0 & -2 \end{pmatrix}
\end{eqnarray}
These Hermitian and traceless matrices obey the trace orthonormality condition $\text{Tr}[G_i G_j] = 2\delta_{ij}$. Along with the identity $G_0 = \mathbb I$, the standard Gell-Mann matrices form an operator basis for $SU(3)$ and can therefore generate any unitary through exponentiation or represent a density matrix as $\rho = \frac{\mathbb I}{3} + \sum^{8}_{i=1} b_i G_i $. Similar to the PTM, we now define a GTM for a process on the three-level system with its elements given by
\begin{equation}
\label{eq:gtm}
    \Lambda_{ij} = \frac{1}{d}\text{Tr}[G_i \mathcal E(G_j)].
\end{equation}

In $SU(d)$ we can therefore construct a GTM using a set of Hermitian and  traceless matrices that can generate any unitary in a $d$-dimensional Hilbert space. These generalized Gell-Mann matrices~\cite{bertlmann2008bloch} are defined as the following $\frac{d(d-1)}{2}$ symmetric matrices
\begin{equation}
    G_{\text{sym}} = \ketbra{k}{l} +  \ketbra{l}{k} ;\,\,\,\,  1\le k \le l \le d,
\end{equation}
$\frac{d(d-1)}{2}$ anti-symmetric, complex-valued matrices
\begin{equation}
    G_{\text{antisym}} = -i \ketbra{k}{l} + i \ketbra{l}{k} ; \,\,\,\,  1\le l \le k \le d,
\end{equation}
and $(d-1)$ diagonal matrices
\begin{equation}
    G_{\text{diag}} = \sqrt{\frac{2}{m(m+1)}}\mleft(\sum^m_{j=1}\ketbra{j}{j}  - m \ketbra{m + 1}{m + 1}\mright) ;\,\,\,\,  1\le m \le d - 1.
\end{equation}
These $d^2 - 1$ matrices $G_1, G_2, ..., G_{d^2 - 1}$, together with the d-dimensional identity $\mathbb I_d$, generalize the idea of using the standard Gell-Mann matrices to represent operators in $SU(d)$ and we use them to construct the GTM. We have used the normalization such that $\text{Tr}[G_i G_j] = 2\delta_{ij}$.

We now show how we adapt the idea of the generalized Gell-Mann matrices to visualize a process in the logically encoded space of a bosonic system with $d=32$. The GTM shows how the different operator basis elements are transformed by $\mathcal E$. Similar to the qutrit system with $d=3$, we can construct the generalized Gell-Mann matrices for $d$-dimensional Hilbert spaces and evaluate \eqref{eq:gtm} for the process $\mathcal E(\rho) = \sum_i^r K_i \rho K_i^{\dagger}$.

The generalized Gell-Mann matrices in the standard form are written using a set of $d$ orthonormal vectors $\ket v_0, \ket v_1, ..., \ket v_d$. We are interested in characterizing logical operations (such as the $X$ gate) in a specific bosonic encoding. We therefore select the basis vectors to include the logical states for our encoding $\ket{v_0} = \ket{0_L} = \ket{2}$, $\ket{v_1} = \ket{1_L} = \frac{\ket{0} + \ket{4}}{\sqrt{2}}$ and complete the basis elements accordingly up to $d=32$ as 
\begin{equation}
\ket{0_L}, \ket{1_L}, \tfrac{\ket{0} - \ket{4}}{\sqrt{2}}, \ket{1}, \ket{3}, \ket{5}, \ket{6}, \ket{7}, ..., \ket{d}.
\end{equation}

Any Gell-Mann matrix $G_i$ can be written as
\begin{equation}
G_i = \sum_{nm} (G_i)_{nm} \ketbra{v_n}{v_m}.
\end{equation}
where any $\ketbra{v_n}{v_m}$ can be expressed using our modified Fock basis that includes the encoded states.

In case of $d = 32$, we have a total of $1023$ Gell-Mann matrices which, combined with the identity, gives a $1024 \times 1024$ GTM. However we are only interested in the first block of elements in the top left corner corresponding to the logical states. Additionally, we are also interested in those elements of the GTM that couple the logical states and possible error states outside the encoded subspace of logical states, e.g., with higher-level Fock states. We therefore sort the Gell-Mann matrices after setting $(\mathbb{I}_d, G_1, G_2, G_3)$ to ($\mathbb{I}_d, X, Y, Z)$ where $(X, Y, Z)$ represent Pauli-like operators for the full system as
\begin{equation}
X = \begin{bmatrix} 
    \sigma_x & 0 & \dots \\
    \vdots & \ddots & \\
    0 &        & 0
    \end{bmatrix}
    ,\quad 
Y = \begin{bmatrix} 
    \sigma_y & 0 & \dots \\
    \vdots & \ddots & \\
    0 &        & 0
    \end{bmatrix}
    ,\quad
Z = \begin{bmatrix} 
    \sigma_z & 0 & \dots \\
    \vdots & \ddots & \\
    0 &        & 0
    \end{bmatrix} .
\end{equation}

In this rewriting of the Gell-Mann matrices, we can now see the similarity with the standard Pauli transfer matrix for operations in the logically encoded space of bosonic states. As an example, for the logical $X$-gate, we expect the top-left block of the GTM to be a diagonal matrix with entries $1, 1, -1, -1$.
\newpage
\section{Error budget}
\begin{figure}[h]
    \centering
    \includegraphics[width=0.6\columnwidth]{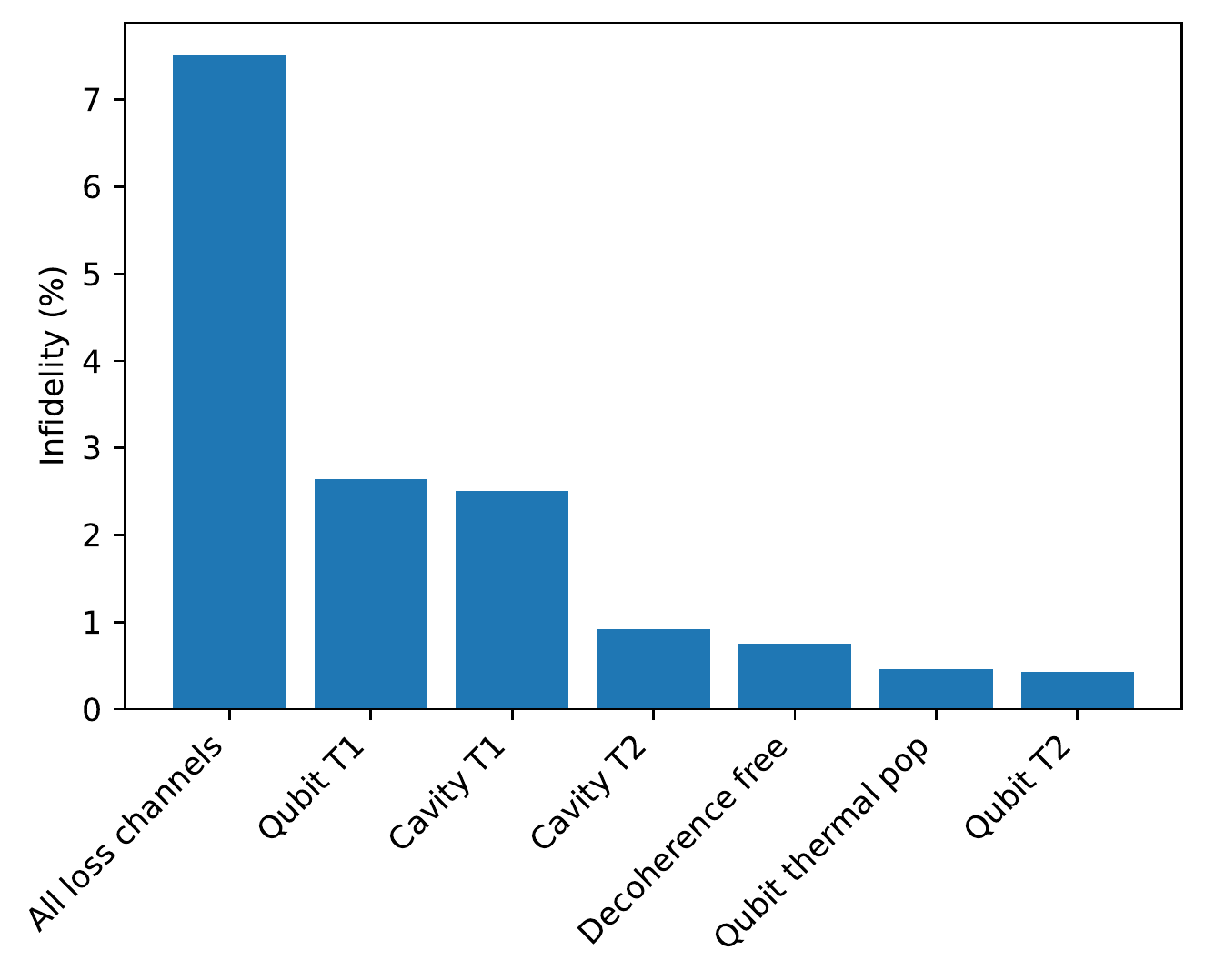}
    \caption{Simulated error budget for the usual loss channels. The case 'Qubit thermal pop' starts the simulation with the qubit in the corresponding thermal state.}
    \label{fig:error_budget}
\end{figure}

The error budget is calculated by simulating how the optimized pulses act on the cavity and ancilla qubit. The decoherence free infidelity arise due to the finite number of SNAP-displacement blocks, which are unable to realize a perfect gate. By only activating a single decoherence mechanism at the time, we simulate and address the corresponding infidelity contribution to each individual channel by subtracting the decoherence free infidelity. In \figref{fig:error_budget} we see that qubit and cavity T1 times give about equal contribution to the infidelity, even though their rates have an order of magnitude difference. 

\section{Limitations of the encoder-decoder process tomography protocol}
A common implementation of quantum process tomography of a logical qubit encoded in a bosonic mode is the encode-decode protocol~\cite{heeres2017implementing}. In the protocol, an ancilla qubit is coupled to a bosonic cavity mode, i.e., the system lives in the joint Hilbert space $\hat H = \hat H_{anc} \otimes \hat H_{cav}.$ The computational space of the logical qubit is spanned by $|0_L\rangle$ and $|1_L\rangle$, whereas all other orthogonal states in the cavity constitute the error space. Process tomography on the encoded qubit is done by first preparing the ancilla qubit in some input state, then encoding (mapping) that state onto the encoded qubit, then executing a gate with the aim to manipulate the logical information, and finally decoding (mapping) the logical information onto the ancilla qubit. Standard two-level tomography of the ancilla qubit is then performed. 

In the following, we will assume a perfect encoder and decoder, and that the logical gate does not entangle the cavity and the ancilla qubit. That is, the ancilla-cavity state is separable $\rho = \rho_{anc}\otimes\rho_{cav}$. The purpose of the decoder is to map the cavity state (when it is in the computational subspace) to the ancilla two-level system. Here, we will study the 'minimalistic decoder', where leakage errors translate to measuring the ancilla preparation state.
The minimalistic decoder swaps only the logical states to the ancilla according to
\begin{equation}
    D_{min} = |e\rangle|0_L\rangle\langle 1_L|\langle g| + |g\rangle|1_L\rangle\langle 0_L|\langle e| + |g\rangle|0_L\rangle\langle 0_L|\langle g| + |e\rangle|1_L\rangle\langle 1_L|\langle e| +  \sum_k \Bigl(|g\rangle|\psi_k\rangle\langle \psi_k|\langle g|+|e\rangle|\psi_k\rangle\langle \psi_k|\langle e|\Bigr)
\end{equation}
while 'identity' acts on the remaining states $\langle\psi_k|\psi_{k'}\rangle = \delta_{k,k'}$ and $\langle\psi_k|i_L\rangle = 0$ for $i=0,1$. For this type of decoder, as soon as we leak out to the error states, the 'identity' operating on these states result in the initial state of the ancilla leaking into our measurement result
\begin{equation}
    \rho_{meas}=\textrm{Tr}_{cav}\left[D_{min}\left(\rho_{anc}\otimes|\psi_k\rangle\langle\psi_k|\right)\right] = \textrm{Tr}_{cav}\left[\rho_{anc}\otimes|\psi_k\rangle\langle\psi_k|\right] = \rho_{anc}.
\end{equation}
Hence, for the minimalistic decoder, we have maximum leakage of information from the ancilla initial state into our measured results.

\begin{figure}[h]
    \centering
    \includegraphics[width=1\columnwidth]{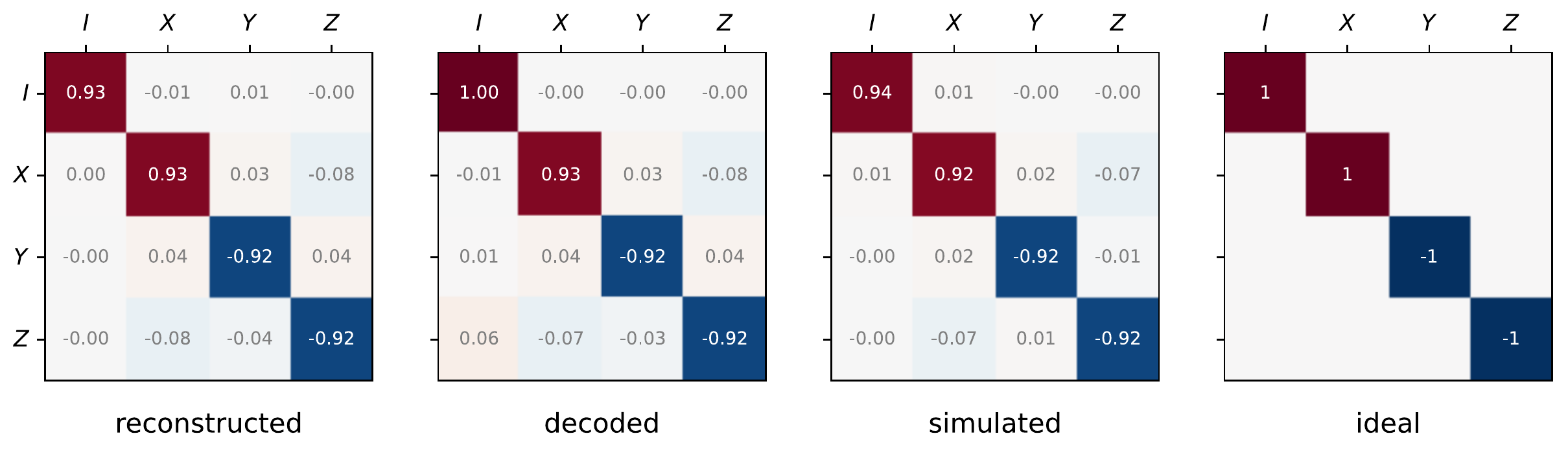}
    \caption{Comparison between the reconstructed, decoded, simulated and ideal Pauli transfer matrices. The matrices present how the process acts on the Pauli basis of the logical subspace. For the reconstructed and simulated matrices, the Pauli basis lives inside a larger Hilbert space.}
    \label{fig:dec}
\end{figure}

To see the effect of the decoder, we take the full reconstructed process $\varepsilon_{rec}$ and apply it to the ideal cardinal points of the encoded qubit to obtain the output states. We then decode the output states by $D_{min}$ and trace out the cavity. In~\figref{fig:dec} we see that since the decode protocol cannot distinguish leakage outside of the computational subspace, it erroneously shows the trace preservation in the logical subspace. In addition, the decode protocol is predicting more $\sigma_z$ (bottom-left element) no matter what the input state is, which corresponds to measuring the ancilla initial ground state. Therefore, it is a sign of a leakage error of this specific decoder, however, without having access to the full Gell-Mann transfer matrix, it is not possible to correctly diagnose the error.

In conclusion, the encoder-decoder protocol can generate close to correct results for high-fidelity processes. However, as soon as we have leakage errors which start to occupy error states in the cavity, the information from the encoder-decoder protocol starts to depend on the exact implementation of the encoder. We showed an example of a minimalistic decoder but in general the decoder can be a black box when optimal-control strategies are used. Hence, in order to draw conclusions from the encoder-decoder protocol, full characterization of the decoder process should generally be performed, which is as complex as directly characterizing the full target process itself. On the contrary, the coherent-state quantum process tomography method described in this work can  correctly and transparently unravel all the errors discussed here.


\end{document}